\documentclass[12pt]{article}
\usepackage[latin1]{inputenc}
\usepackage[english]{babel}
\usepackage{babel} 
\usepackage{graphics}
\usepackage{setspace}

\makeatletter
\newcommand{\LyX}{L\kern-.1667em\lower.25em\hbox{Y}\kern-.125emX\spacefactor1000}

\let\SF@@footnote\footnote
\def\footnote{\ifx\protect\@typeset@protect
    \expandafter\SF@@footnote
  \else
    \expandafter\SF@gobble@opt
  \fi
}
\expandafter
\def\csname SF@gobble@opt \endcsname{\@ifnextchar[  \SF@gobble@twobracket
  \@gobble
}
\edef\SF@gobble@opt{\noexpand\protect
  \expandafter\noexpand\csname SF@gobble@opt \endcsname}
\def\SF@gobble@twobracket[#1]#2{}
\makeatother
\input tcilatex
\QQQ{Language}{
American English
}

\begin{document}

\title{\Huge Theoretical Considerations of Electrochemical Phase Formation in a Frank-van der Merwe System}
\author{ 
M.C. Gim\'{e}nez, M.G. Del P\'{o}polo, E.P.M. Leiva \thanks{Corresponding author. Fax 54-351-4334189; 
e-mail: eleiva@fcq.unc.edu.ar}
\\
Unidad de Matem\'atica y F\'{\i}sica, Facultad de Ciencias Qu\'{\i}micas,\\
Universidad Nacional de C\'{o}rdoba, 5000\\
C\'{o}rdoba, Argentina\\
\\
\and S.G. Garc\'{\i}a, D.R. Salinas, C.E. Mayer
\\
Instituto de Ingenier\'{\i}a Electroqu\'{\i}mica y Corrosi\'on,\\
Departamento de Qu\'{\i}mica e Ingenier\'{\i}a Qu\'{\i}mica,\\
Universidad Nacional del Sur, Bah\'{\i}a Blanca, Buenos Aires, Argentina\\
\\
\and W.J. Lorenz
\\
Institute of Nanotechnology,
Universit\"at Karlsruhe\\
Kaiserstr. 12, D-76131 Karlsruhe, Germany.
\\
}
\maketitle

\newpage
\begin{abstract}
Static calculation and preliminary kinetic Monte Carlo simulation studies
are undertaken for the nucleation and growth on a model system which follows a Frank-van der
Merwe mechanism. In the present case, we consider the deposition of Ag on Au(100) and
Au(111) surfaces. The interactions were calculated using the embedded atom
model. The process of formation and growth of 2D Ag structures on Au(100) and Au(111)
is investigated and the influence of surface steps on this phenomenon is
studied. Very different time scales are predicted for Ag diffusion on
Au(100) and Au(111), thus rendering very different regimes for the
nucleation and growth of the related 2D phases. These observations are
drawn from application of a model free of any adjustable parameter.\bigskip%
\ 

\end{abstract}

\newpage

\section{Introduction}
The electrodeposition of a metal (Me) onto a foreign solid surface (S)
is one of the most extensively studied subjects in surface electrochemistry. 
This electrochemical phase formation phenomenon is a key aspect in important 
technological processes such as electroplating and electrocatalysis. In the 
last years, the application of \textit{in-situ} local probe microscopy (SPM) techniques 
such as scanning tunneling microscopy (STM) and atomic force microscopy (AFM) 
provided a powerful tool to observe the initial stages of these processes on an 
atomic level. In a number of cases, metal overlayers can be electrodeposited 
onto a foreing metal substrate at a potential that is less negative than 
the Nernst equilibrium potential of the 3D metal phase. This so-called 
underpotential deposition process (UPD) occurs in the undersaturation or 
underpotential range, given by convention as:
\begin{equation}
\Delta E = E -E_{3DMe} > 0 \label{1}
\end{equation}
where $E$ is the actual electrode potential, $E_{3DMe}$ represents the equilibrium
potential of the 3D metal phase, and $\Delta E$ is the underpotential shift.

The UPD process has been well characterized for many systems with the SPM 
techniques and the preceding formation of metal phases of low dimensionality 
(0D, 1D, 2D) have been analyzed \cite{Lorenz_Libro,L2,L3,L4,L5,L6}. The stability ranges of these \textit{i}D Me 
phases (\textit{i} = 0, 1, 2, 3) can be formally described by Nernst-type equations: 
\begin{equation}
E_{iDMe} = E_{iDMe}^{0} + \frac{RT}{zF}\log{\frac{a_{Me^{z+}}}{a_{{iDMe}}}} \label{2}
\end{equation}
where $a_{Me^{z+}}$ denotes the activity of $\textrm{Me}^{z+}$ ions within the electrolyte, $a_{iDMe}$ is the 
activity of the \textit{i}D Me phase, which is a constant for a condensed Me phase,
and $E_{iDMe}^{0}$ represents the corresponding equilibrium potential. Usually, the 
electrochemical formation of the low dimensional condensed phases occurs 
under supersaturation conditions, whereas their dissolution takes place under 
undersaturation conditions. For an \textit{i}D Me phase, the corresponding overpotential 
deposition (OPD) or supersaturation range, is given by:
\begin{equation}
\eta_{iDMe}=E-E_{iDMe}<0 \label{3}
\end{equation}
where $\eta_{iDMe}$ is the overpotential corresponding to the \textit{i}D Me phase.
The formation of these \textit{i}D Me phases is strongly influenced by surface defects 
like kinks, vacancies, chemical impurities, monatomic steps, stacking faults, 
etc. The crystal surface can also be considered as a 2D crystal imperfection 
and plays an important role in the UPD phenomena. Thus, the stepwise formation 
of low-dimensional Me phases on a substrate in the underpotential range are 
generally characterized by the formation of 0D and 1D phases at relatively 
high underpotentials, followed by the formation of expanded and condensed 2D phases on atomically 
flat terraces at lower underpotentials.

In  a previous work \cite{ceci}, we considered
a theoretical model suitable to analyze the existence of these different phases
when a metal is deposited on a foreign substrate with defects. We employed a lattice 
model along with a grand Canonical Monte Carlo simulation to study the thermodynamic
features of the deposition process. One of the novel features of this model consisted in
the realistic modelling of the interactions between the particles of the system, which was made 
using the embedded atom method (EAM). In this way, we introduced in a simple lattice model the many body 
features of the metallic binding which are indispensable for a reliable description of the 
metal-metal interactions. In agreement with the expectation drawn from the experience with 
experimental systems, we observed the correct sequence of appeareance of 0D,1D, and 2D 
phases when changing the chemical potential of the system in a direction corresponding to 
polarization of the surface towards negative potentials. Furthermore, the different regions
of existence of the phases were separated by chemical potentials  of a few
hundreds of meV, whose values are of the order of magnitud of those observed experimentally.

In some cases, the 2D Me phase formation is not reached after sequence of appearance of
0D and 1D phases, but proceeds by the formation of 2D expanded commensurate Me adlayers which are 
transformed at lower undersaturation into condensed metal monolayers via a first 
order phase transition. Such phase transitions are observed, for example, during the Ag UPD 
on both Au(111) and Au(100) surfaces \cite{SGarcia,Ag/Au_Nuevo}.

In summary, UPD systems are very complex ones, as they involve phases of different dimensionalities
which are mainly characterized by:
\begin{enumerate}
\item Specific adsorption of metal ions
\item Electronic charge transfer
\item Metal adatom-adatom and substrate-adatom interaction, which cannot be 
rigorously disentangled due to the non-additive property of the metal-metal
interaction.
\item Undersaturation and/or supersaturation conditions with respect to the several
(\textit{i}D) phases that may appear on the system.
\item Crystallographic misfit, responsible in many cases for the lack of registry
between the adsorbate and the substrate.
\item Cosorption and competitive sorption phenomena of anions as well oxidic 
species leading also to the formation of low-dimensional anion or oxidic systems 
in the corresponding undersaturation ranges.
\item The influence of the zero charge potential of the substrate
on the electric field conditions at the interface and on adsorption and phase
formation phenomena.
\item Surface site exchange processes leading to the formation of \textit{i}D surface Me-S alloys.  
\end{enumerate}

In principle, a general description of the UPD phenomena would require a model
incorporating all these items, something that is possible in principle with the current status
of the developing of computers but still a formidable
task for the small community of theoretical electrochemistry. However, more 
simple models may be formulated, including some of the points mentioned above,
and a number of properties may be inferred for those systems whose salient features 
correspond to the model selected.

The key features of the present work concentrate on a reliable description of
the metal-metal interaction by the EAM, and a lattice model which allows the 
consideration of relative large systems. Thus we can make the following comments 
in relation with the points mentioned above:
Points 1), 2), 6) and 7) cannot be directly addressed in the present model, 
since this would require a detailed description of the metal/solution interface,
that is only possible using first-principles calculations. Therefore, we 
shall only deal with the deposition of atoms at different rates, without being
concerned on how these rates are related to the corresponding potential changes
at the interface.
Point 3) is one of the strenghts of the present formulation, since the 
interactions in the EAM have been devised to account for the properties
of the metals involved and their alloys. With respect to
point 4), in the present work we concentrate on the deposition process, 
neglecting adatom desorption. So we set supersaturation conditions for the
formation of the several low dimensional Me phases.
Point 5) is a very complicated one, since the crystallographic misfit 
induces in many cases the existence of incommensurate structures,
which cannot be addressed by lattice models. Thus, the way to overcome 
this problem is by choosing a system with a negligible 
crystallographic misfit. In this respect, the deposition of Ag on Au appears as 
an optimal candidate.
Concerning point 8), site exchange with the substrate could be considered within
 the present model, but preliminary
studies by molecular dynamics show that this effect is not of primary
importance in the submonolayer range.

According to the previous discussion, we perform in this 
work model calculations for Ag deposition on Au(111) and Au(100) surfaces, modelling the metal-metal
interactions by the embedded atom method. These systems appear as optimal candidates due
to a negligible crystallographic misfit. On the other hand, both Ag and Au atoms are mobile
in our model, so that 2D alloys could naturally appear during the course of the 
simulation but this 2D alloying process has not been observed. Therefore, the aim of this 
work is to set the basis for a systematic study of a system  involving electrochemical phase 
formation in a Frank-van der Merwe system and emphasize the role of metal-metal interactions 
in the surface processes accompanying this phenomenon.

\section{Some experimental facts of the system \\ $\textrm{Au}(hkl)/\textrm{Ag}^+$}

The system $\textrm{Au}(hkl)/\textrm{Ag}^{+}$ is a typical example for Me UPD on a foreign substrate S,
 with strong Me-S interaction but negligible Me-S misfit $(d_{0,Au} = 0,2884\hspace{1mm}nm,\hspace{1mm} 
 d_{0,Ag} = 0,2889 \hspace{1mm}nm)$. Thus, the Frank-van der Merwe or layer by layer growth mechanism is 
 expected to operate in this system. Our previous studies \cite{SGarcia,Ag/Au_Nuevo} have demonstrated 
 that the Ag UPD on Au(111) and Au(100) occurs in the underpotential range 
 $0 \leq \Delta E/\textrm{mV} \leq 720$. For the system $\textrm{Au(100)/Ag}^{+}$, the formation of an expanded 
 $\textrm{Au(100)}-c(\sqrt{2}\times 5\sqrt{2})\textrm{R45°Ag}$ overlayer is observed within the potential range 
 $200 \leq \Delta E/\textrm{mV} \leq 550$.  At lower underpotentials a limited growth of steps 
 occurs and 2D Ag islands are formed involving simultaneous 2D nucleation 
 and growth at steps and flat terraces (Figure 1). The atomic structure on 
 top of islands as well as terraces is quadratic with an interatomic distance 
 of $d_{0,Ag} = 0.29 ± 0.01 \hspace{1mm}\textrm{nm}$. These morphological and atomic observations 
 indicate that the expanded structure transforms into a $\textrm{Au(100)}-(1\times1)\textrm{Ag}$ 
 phase via a first order phase transition. The steps of the 2D Ag islands 
 grow slowly depending on the potential, until a complete Ag UPD monolayer 
 is formed at $\Delta E < 15 \hspace{1mm}\textrm{mV}$. On the other hand, in the system 
 $\textrm{Au(111)/Ag}^{+}$ the experiments showed an expanded $\textrm{Au(111)}-(4\times4)\textrm{Ag}$ structure in the range 
 $50 \leq \Delta E/\textrm{mV} \leq 500$ which is transformed to a $\textrm{Au(111)}-(1\times1)\textrm{Ag}$ phase at lower 
 underpotentials. The interatomic distance observed $(d_{0,Ag} = 0.28 ± 0.01\hspace{1mm}\textrm{nm})$ 
 indicates the formation of an hexagonal close-packed Ag monolayer. It is 
 important to note that, in this case, the formation of the condensed 2D Ag 
 phase on Au(111) occurs preferentially at monatomic steps of the substrate, 
 without the formation of 2D islands (Figure 2). The atomic structures and 
 morphologies described above were observed in both sulphate and perchlorate 
 solutions. In addition, electrosorption valency measurements have indicated 
 that coadsorption or competitive adsorption of sulphate or perchlorate anions 
 can be excluded. Nevertheless, as was suggested \cite{SGarcia,Ag/Au_Nuevo}, a nearly constant anion 
 coverage in the entire Ag UPD range cannot be excluded because Ag UPD on 
 $\textrm{Au}(hkl)$ occurs at positive potentials with respect to the potential of zero 
 charge of $\textrm{Au}(hkl)$ and $\textrm{Ag}(hkl)$.

 From this brief experimental outlook, it is clear that the model we are 
 presenting in this work is still missing an important number of features to 
 reflect the experimental situation. However, as we shall see below several
 aspects of the experimental system coincide with the predictions of the model.

\section{Model and simulation method}

\subsection{Interatomic potential}

The choice of proper interatomic potentials to perform the simulations is
one of the key elements of the model. Several methods are available to
calculate the total energy of a many-particle metallic system, with a
computational effort comparable to that of a pair potential \cite{Carlsson}.
It is worth mentioning the embedded atom method (EAM) \cite{Daw-Baskes}, the
N-body potentials of Finnis and Sinclair \cite{Finnis}, the second-moment
approximation or Tight-binding (TB) \cite{Carlsson} and the glue model (GM) 
\cite{Ercolessi_2}. In this work we employ the embedded atom method \cite
{Daw-Baskes} because it reproduces important characteristics of the metallic
binding that cannot be obtained using simple pair potentials.

The EAM considers that the total energy $U_{tot}$ of an arrangement of $N$
particles may be calculated as the sum of energies $U_i$ corresponding to
individual particles 
\begin{equation}
U_{tot}=\sum_{i=1}^NU_i  \label{4}
\end{equation}
where $U_i$ is given by 
\begin{equation}
U_i=F_i(\rho _{h,i})+\frac 12\sum_{j\neq i}V_{ij}(r_{ij})  \label{5}
\end{equation}
$F_i$ is denominated embedding function and represents the energy necessary
to embed atom $i$ in the electronic density $\rho _{h,i}$ at the site at
which this atom is located. $\rho _{h,i}$ is calculated as the superposition
of the individual electronic densities $\rho _i(r_{ij}):$%
\begin{equation}
\rho _{h,i}=\sum_{j\neq i}\rho _i(r_{ij})  \label{6}
\end{equation}
Thus, the attractive contribution to the EAM potential is given by the
embedding energy, which accounts for many-body effects. On the other hand,
the repulsion between ion cores is represented through a pair potential $%
V_{ij}(r_{ij})$, which depends only on the distance between the cores $%
r_{ij} $: 
\begin{equation}
V_{ij}=\frac{Z_i(r_{ij})Z_j(r_{ij})}{r_{ij}}  \label{7}
\end{equation}
where $Z_i(r_{ij})$ may be envisaged as a sort of effective charge,
dependent on the nature of the particle $i$.

The EAM has been parametrized to fit experimental data like elastic
constants, dissolution enthalpies of binary alloys, bulk lattice constants
and sublimation heaths \cite{Daw-Baskes}. Pair functionals have been
successfully used for surface diffusion studies and adsorption of metals on
metallic surfaces \cite{Votter, Haftel, Cu/Ag}.

\subsection{Lattice model}

Lattice models for computer simulations have been widely used in studies of
nucleation and growth, because they allow simulations with a large number of
particles at a relatively low computational cost. The reason for this advantage is the use
of fixed rigid lattices that restrict enormously the number of possible
configurations for the adsorbate as compared with a model where in principle
all positions in space are available. While continuum Hamiltonians are much
more realistic in those cases where epitaxial growth of an adsorbate leads
to incommensurate adsorbed phases \cite{Pb/Ag} or to adsorbates with large
coincidence cells, the use of lattice models may be justified on the basis
of experimental evidence or continuum computer simulations that predict a
proper fixed lattice geometry. In the present case, we have strong evidence
from simulations within the canonical Monte Carlo method \cite{Portugal}
which indicates that at least one of the phases occurring during Ag
underpotential deposition on Au(111) and Au(100) possess a pseudomorphic
structure. In fact, our continuum MC simulations showed that Ag monolayers
adsorbed on Au(111) and Au(100) spontaneously acquired a $(1\times 1)$
coincidence cell in agreement with the experimental finding at low
underpotentials. For this reason, we shall employ here a lattice model to
represent the square (100) and the hexagonal (111) surface lattices in
kinetic Monte Carlo simulations. Square lattices and hexagonal lattices of
different sizes with periodical boundary conditions are used in the present
work to represent the surface of the electrode. Each lattice node represents
an adsorption site for a Ag or a Au atom.

Two further approximations in the present model must also be mentioned.
First, we neglect the effect of the presence of solvent molecules. This
approximation should not be critical as long as the partial charge on the
adatoms is small, thus minimizing the ion-dipole interactions. Second, we
also neglect effects related to specific anion adsorption. This may not be
true for the experimental system, but thin layer twin electrode experiments
gave no indication of any change in the amount of anion adsorption upon
building the adsorbed monolayer \cite{Ag/Au_Nuevo}. On the other hand, the
possibility of a nearly constant anion layer on the bare substrate Au(100)
as well as on a Ag UPD modified Au(100) has been indicated \cite{Ag/Au_Nuevo}%
. In the case of other systems, the adatom-anion interactions have been
shown to be very important \cite{Aniones} in playing a decisive role for
determining the energetics of the system. This has been recently analyzed by
some of us in thermodynamic terms \cite{upd-eam}.

\subsection{Calculation of Gibbs energy of cluster formation.}

The initial stages of a phase formation are of primary importance, because
the competition of different processes occurring during this phenomenon
determine in many practical cases the final structure of the deposit. The
thermodynamic foundations of the initial stages of bulk phase formation are
well settled and have been thoroughly discussed in advanced books on the
field \cite{Lorenz_Libro}. For this reason, we shall address here briefly
the topics relevant for the present calculations. The free energy change $%
\Delta G(N)$ produced when a three-dimensional cluster of $N$ atoms is
formed on a surface at a certain overpotential $\eta_{3DMe} $ can be written as:
\begin{equation}
\Delta G(N)=-Nze\left| \eta_{3DMe} \right| +\Phi (N)  \label{deltaGnuc}
\end{equation}
where $z$ is the valence of the deposited ions, $e$ is the elementary charge
and $\Phi (N)$ is the energy consumed for the formation of the new interface
boundaries. In the 3D case, this quantity is related to the specific
surface energies $\sigma _i$ of the facets of the cluster, the specific
interface energy $\sigma _{j^{*}}$ and the surface energy of the substrate
according to:
\begin{equation}
\Phi (N)=\sum \sigma _iA_i+A_{j^{*}}(\sigma _{j^{*}}-\sigma _{sub}) \label{fin}
\end{equation}
where $A_i$ denotes the respective surface areas. $\Phi (N)$ can also be
calculated within the atomistic model, as long suitable potentials for the
system are available. If we called $E_{c+S}$ the energy of a system
consisting of a cluster of $N$ atoms of the metal Me on a substrate S, $%
E_S$ the energy of the free substrate, and $E_b$ the binding energy per atom
of the bulk adsorbate metal, $\Phi (N)$ would be given by: 
\begin{equation}
\Phi (N)=E_{c+S}-E_S-N\ E_b  \label{fin2}
\end{equation}
If the energy of the atoms in the cluster was equal to that in the bulk,
Eq.\ref{fin2} would yield $\Phi (N)=0$, and a vanishing small
overpotential would be enough for the growth of the new phase according to
Eq. \ref{deltaGnuc}.

In case of 2D nucleation, which is the current subject of our interest, an
equation similar to (\ref{deltaGnuc}) is valid. However, instead of defining
the overpotential with respect to the deposition of bulk Me, it is more
suitable to refer to it as the thermodynamic deposition potential of the
monolayer. Thus, the excess energy $\Phi _{2D}(N)$ will be calculated
from:
\begin{equation}
\Phi _{2D}(N)=E_{c+S}-E_S-N\ E_{mon}^{Me} 
\end{equation}
where $E_{mon}^{Me}$ is the binding energy of the \thinspace Me atoms in the
monolayer. In consequence, $\Phi _{\textrm{2D}}(N)$ takes into account the excess
energy connected with the occurrence of the border of the 2D islands.

For a given number of atoms $N$, several values of $\Phi _{2D}(N)$ can be found
depending on the geometry of the 2D cluster. For each $N$ we have only
considered the energy corresponding to most stable configuration. This was
obtained by means of a simulated annealing procedure. This technique has
often been used to obtain minimal energy structures or to solve ergodicity
problems. A suitable way to implement this technique is through the canonical Monte
Carlo method at different temperatures. In other words, a given number of atoms, $N$,
is selected, and a simulation allowing the motion of the atoms on the
surface is started at a high initial temperature $T_o$, in the order of
10$^3$ K. The system is later cooled down following a logarithmic law:
\begin{equation}
T_f=T_oK^{N_{cycles}}  \label{14}
\end{equation}
where $T_f$ is the final temperature, $N_{cycles}$ is the number of cooling
steps and $K$ is a positive constant lower than 1. A few hundreds of
thousands of Monte Carlo steps are run at each temperature in order to allow
an extensive exploration of the configuration space and the simulation stops
when $T_f$ is reached.

\subsection{Model for Dynamic Monte Carlo simulations}

Although Monte Carlo methods have been traditionally related to the study of
equilibrium properties, it is possible, if some conditions are fulfilled, to
use them to compute the time evolution of a system. The foundations of
dynamical Monte Carlo simulations have been discussed by Fichtorn and
Weinberg \cite{DMC} in terms of the theory of Poisson processes. According to
this method, the sampling of the system must implicate transition
probabilities based on a reasonable dynamic model of the physical phenomena
involved. Besides fulfilling the detailed balance criterion, the transition
probabilities should reflect a ''dynamic hierarchy '' related to the
processes taking place in the system.

In this preliminary study we are mainly interested in nucleation and growth
phenomena, therefore, we shall consider only two types of processes related to the
growth of the new phase: adsorption of an adatom on the surface and its
diffusion in different environments. Since we are neglecting the desorption
processes, the results of our model will be valid for relatively large
overpotentials for the deposition of the Ag atoms.

To illustrate the model used for the dynamic calculations, we show in Figure
3 a-c the diffusion of a Ag atom on a Au(100) surface in three different
environments. In Figure 3a, a Ag adatom jumps between two equivalent sites
without any near neighbors. In Figure 3b, it has one Ag neighboring atom in
the initial position and none in the final state and in figure 3c the jump
is again between two equivalent positions with one Au nearest neighbor. The
corresponding potential energy curves calculated by the EAM at each stage of
the diffusion are shown in Figure 3d. From these curves we obtain the two
important quantities required to perform the dynamic MC simulation: the
attempt frequency for overcoming the diffusion barrier, calculated for the
curvature of the potential energy surface at the initial state, and the
activation energy, by taking the difference between the energy at the saddle
point and the corresponding value at the initial minimum. These curves were
constructed for \textit{all} possible environments and involved 729 and 6561
energy curves in the case of the (100) and (111) faces respectively.

In the case of the adsorption rates, we assumed the same rate for all free
sites. A more realistic calculation should also take into account different
rates on different environments. This would require a complete knowledge of
the degree of solvation of the discharging ion and detailed information of
the different transition states in different environments. This information
is not available for the present system but could be easily introduced in
our modelling. Experimental evidence show that electrodeposition of metal ions on native or
foreign substrates take place preferentially at 0D and 1D surface
inhomogeneities such as kink and monoatomic steps. In the case of foreign
substrates, 0D metal clusters are preferentially formed at these
inhomogeneities.We have addressed the role of different types of defects by
means of \textit{thermodynamic} Monte Carlo studies in previous work,
confirming this expectation. In the present work, we study the deposition
phenomena from a \textit{kinetic} point of view, analyzing the influence of
the deposition rate on the formation and growth of the 2-D phase on a
foreign surface with a relatively small number of defects.

\section{Results and discussion}

\subsection{Static calculations}

\subsubsection{2D Ag cluster formation on Au(100) and Au(111) surfaces.}

Figure 4 shows the Gibbs energy of 2D Ag cluster formation on Au(100) as a
function of size $N$, calculated according to the method described in section
3.3. Since we are using an atomistic model, our curves are discontinuous.
For this reason, we have fitted our curve for $\eta_{2DAg} =0$ by means of a square
root law, and later employed it to draw the corresponding continuous lines
for $\left| \eta_{2DAg} \right| >0$. With this information we were able to
calculate the size of the 2D critical clusters for different overvoltages,
as reported in Table 1a. The sizes of the critical clusters are relatively
large as compared with the values estimated for nucleation of Ag on a
Ag(100) quasi perfect surface \cite{Lorenz_Libro}, given in Table 2b. In
other words, the overpotentials for nucleation are considerably larger when
Ag nucleates on the foreign Au(100) surface.

Figure 5 shows the Gibbs energy of a 2D Ag cluster formation on Au(111) at
different overpotentials. It can be noticed that for a given overpotential,
the maximum in the curve appears at larger $N$ than in the case of the
Au(100) surface, that is, the energy for the formation of border atoms from
atoms located in the monolayer is larger for the adsorbate on the (111)
face. This fact can be understood in terms of the relative energy change for
this process. In the case of an adsorbate in the (100) monolayer, it has
eight nearest neighbors: 4 Au substrate atoms plus 4 Ag atoms belonging to
the monolayer. This coordination changes to seven (4 Au + 3 Ag) for a border
atom. In the case of the (111) adsorbate, the change is from nine (3 Au + 6
Ag) to seven (3 Au + 4 Ag) at the border, that is, the change of
coordination to produce a border atom from the monolayer is more important
in the case of the more compact adsorbate, originating a stronger energy
change. This is somehow the 2D analogous of the qualitative justification
why more open faces exhibit a larger surface energy than the more compact
ones.

\subsubsection{Diffusion of single Ag atoms on Au(100) and Au(111)}

The growth of a 2D phase after nucleation involves a number of elementary
steps that should be considered in a general formulation of the phenomenon.
This has been analyzed in detail in specialized texts on the field\cite
{Lorenz_Libro}, hence we simply enumerate them here:

\begin{enumerate}
\item  Bulk diffusion of the discharging ions.
\item  Adatom electrodeposition and electrodesorption.
\item  Surface diffusion of the discharged adatoms and diffusion of adatoms along the steps.
\item  Direct attachment to steps or kink sites. 
\end{enumerate}

The relative rates of these processes govern the overall rate and it is of
key importance for the modelling of each system to determine the one(s)
ruling the whole process. For example, in the case of the growth of a 2D Ag
cluster on Ag(100) quasi perfect faces, step 2 is very fast under
the usual deposition conditions yielding an unusually high exchange current
density of Ag atoms. In this way, the average random surface displacement of
adatoms during their stay at the surface is small and surface diffusion
plays only a subordinate role.

The high exchange current density of Ag atoms has also dramatic effects during 
the deposition of the second Ag monolayer on Au(111),
producing very noisy STM pictures under these conditions. On the
contrary, images obtained for the growth of Ag islands on Au(100) are stable
and clear \cite{SGarcia} (Figure 1b, 1c), denoting that the exchange current for Ag atoms on
Au surfaces is considerably lower. On these grounds, we assume in our
simulations and the following considerations that atom deposition occurs
under conditions where atom exchange is negligible.

The diffusion coefficients $D_{hkl}$ of a Ag single atom on
the Au surfaces can be calculated from:
\begin{equation}
D_{hkl}=\frac{na^2\nu }4e^{-\frac{E_a}{kT}} 
\end{equation}
where $a$ is the distance between adsorption sites, $\nu $ is the attempt
frequency, $E_a$ is the activation energy for diffusion and $n=4$ or $3$ for
the (100) and (111) faces respectively. We obtained $D_{100}=3.3\times 10^3\AA^2 s^{-1}$ 
and $D_{111}=1.2\times 10^{10} \AA ^2s^{-1}$, indicating that the
diffusion on the (111) face is much faster than the same process on the (100)
face. These figures already give us a very important hint to predict the
type of nucleation and growth behavior in these two faces under the usual
experimental conditions in voltammetry. In fact, let us consider an
experiment where the reduction current density producing adatoms, $i$, is of the order of
1 $\mu A/cm^2,$ and let us assume that the surface presents terraces of the
order of $40\times40$ nm$^2$. The number of atoms originated at the terrace per
unit time will thus be of the order of $10^2$ $s^{-1}$ . We will denote this
number with $\frac 1\tau $. Thus, once an atom is deposited, it has the
chance to diffuse the length:
\begin{equation}
d_{hkl}\simeq \sqrt{2\ D_{hkl}\ \tau }
\end{equation}
before any other atom is deposited on the average on the same terrace. If we
replace in this equation the $D_{hkl}$ obtained, we get $d_{100}\approx 8$
\AA\ and $d_{111}\approx 1.5\times 10^4$\AA\ respectively. Thus, while on
the Au(100) face the deposited Ag atoms may meet other atoms for the
nucleation and growth of the 2D phase, the Ag atoms deposited on the
Au(111) surface will diffuse unhindered to the border of the terraces and
produce the 2D growth there. Since the diffusion barriers are usually higher 
on (100) surfaces than on (111) ones, we expect that this should be a rather
general result, also valid for other upd systems. It is worth mentioning 
that, as shown in section 2, the experimental Ag/Au(hkl) upd system presents 
a similar behavior, in that the growth of the 2D phase occurs at the border of 
steps on the (111) face, and in the form of islands on the (100) face, besides 
a limited growth at steps.
 
\subsection{Dynamic Monte Carlo simulations.}

\subsubsection{Ag on Au(100)}

We discuss here the results obtained with the simulation 2D box shown in
Figure 6, which is relatively small as compared with real terraces, but
should yield the correct qualitative features. We consider 1600 adsorption
sites distributed over a $115 \times 115\ \AA ^2$ square lattice, delimited
by two rows of monoatomic high Au steps on each side. Periodic boundary
conditions were applied in the direction parallel to the borders. 
Real surfaces present a more complex topology, with  different kinds of 
0D and 1D inhomogeneities like kinks or roughened steps. However, our simulation 
could also represent a situation where kink or steps are already decorated
by 0D or 1D silver phases. 
We only allowed for the deposition of Ag atoms on the terraces at different rates $%
v_{ads}$. We varied $v_{ads}$ between 10$^{-4}$ $s^{-1}$ and 10$^2$ $s^{-1}$
per free adsorption site, corresponding to potentiostatic conditions with an
initial current density ($\Theta \rightarrow 0$) ranging between 2$\times 10^{-2}$ $%
\mu A/cm^2$ and $2\times 10^4\mu A/cm^2$. Although we made several sets of
simulations, we discuss here some representative examples to obtain a
physical insight into the phenomena taking place in each case.

For the slowest rate (10$^{-4}$ $s^{-1}$) we typically observed initially a
single particle diffusing on the surface towards the Au border step and then
remaining attached to the border and diffusing parallel to it. Diffusion
along the borders is relatively fast with respect to surface diffusion,
since $D_{border}^{100}=1.7 \times 10^5$ $\AA^2s^{-1}$. About 15 seconds
later, a second particle entered and diffused to the opposite side, followed
by others that behaved in the same way. Two Ag stripes parallel to the Au
borders grew in this way, being completed at about 650 seconds. After that a
second row started to grow. Diffusion of Ag on the Ag border also showed a
diffusion coefficient of the order of $2\times 10^5$  $\AA^2s^{-1}$. In this
simulation, no Ag island formation was evident on the surface and the growth
occurred at the Au border. The third Ag row was completed at about 1800
seconds.

When the adsorption rate was $v_{ads} = 10^{-3}$ $s^{-1}$, all the particles
deposited up to 50 seconds were found to diffuse to the borders. At that
time, a dimer was formed, which initially diffused on the surface and later
was fixed by addition of more particles. At 170 s a trimer was formed,
giving place to a second island. At 700 seconds, the coalescence started to take
place between the growing islands and the growth at the Au borders. At 7250
seconds, only a single vacancy remained on the surface, diffusing with $%
D_{vacancy}=4.2\times 10^3$ $\AA^2s^{-1}.$

The run with $v_{ads}=$ 10$^{-2}$ $s^{-1}$ presented several interesting
features, since at this deposition rate (corresponding to $\approx 2$ $\mu
A/cm^2$) the number of islands appearing in our simulation box allows for
statistical analysis. For example, the number of islands was followed as a
function of time. In the simulation probed, the number of islands reached a
maximum of 7 at about 15 seconds (Figure 7), decreasing later and generating
islands of vacancies at $t\approx 400\ s.$ At $t=630\ s,$only a vacancy
remained diffusing on the surface and the surface was completely covered by
Ag atoms at $t=860\ s$. In Figure 8 we show snapshots of a simulation with
this adsorption rate.

In the case of deposition rates of $10^{-1}$ $s^{-1}$ and larger, a
coexistence of growing islands and diffusing particles is observed from the
very beginning of the simulation. Typical times for the completion of the
monolayer are reported in Table 2. It can be noticed that they follow the
expected logarithmic law.

Figure 9 shows the number of islands as a function of the coverage degree
for different deposition rates. Each of the curves represents an average
over six simulation runs. It can be noticed there how the number of islands
increases with the deposition rate, with a maximum in the range $0.1<\Theta
<0.2$ which is progressively better defined at the higher rates.

Unfortunately there are no experiments available for the present simulation
conditions, consequently a direct comparison is not possible. STM-voltammetric
experiments where the deposition rate is of the same order of magnitude as
some of our results ($i\approx 2\ \mu A/cm^2$,corresponding to  $v_{ads}=10^{-2}s^{-1})$), 
show the occurrence
of growing islands \cite{SGarcia}, generated at times of the order of a few
seconds, which correspond to the simulation times where the islands appear
in our simulations. It must be stressed, however, that the deposition
history in the experiment is completely different, since the pseudomorphic
phase appears after the voltammetric formation of an expanded 
$\textrm{Au(100)}-c(\sqrt{2}\times 5\sqrt{2})R45^o \textrm{Ag}$ 
phase which is not considered in the
present model.

\subsubsection{Ag on Au(111)}

The simulation box was analogous to that of the Au(111) face, and the
adsorption rates were varied between 10$^{-3}\leq v_{ads}/s^{-1} \leq 10^8$. Since the
activation barriers on this surface are very low, the absolute-rates model
employed to calculate the transitions of atoms on the surface should deliver
only qualitative predictions. Furthermore, some transitions on these
surfaces were not activated at all. We assigned to these processes an
arbitrarily high rate, so they made a negligible contribution to the time
accumulation in the simulation and we were able to follow the relatively
slow processes occurring on the surface.

In the case of the Au(111) surface, the adsorption sites constitute an
hexagonal lattice. Two kinds of adsorption sites occur, usually denoted as
'hcp' and 'fcc', depending on the position of the adsorbing atom with
respect to the second lattice plane of substrate. In our model these sites
have practically the same energy, but the lattices they define cannot be
filled simultaneously due to steric hindrances. Thus, under some simulation
conditions two domains (one fcc and one hcp) of the adsorbate appear, with a
corresponding domain border.

For $v_{ads}\leq 10^3$ $s^{-1}$, all atoms reached the border of the box.
However, since we are dealing now with the (111) face, the borders are no
longer equivalent. One of them exhibits a facet with square symmetry (that
we shall label \{100\}) and the other presents three fold sites \{111\}. In
the case of the \{100\} border, we obtained straight Ag stripes parallel to
the step, while a triangle like decoration was found in the case of the
\{111\} step (Figure 10). This different type of growth occurs in our model on
kinetic ground as a consequence of the fact that only jumps between
nearest neighbors are allowed. Molecular dynamic calculations would be very
helpful to elucidate this point.

For $v_{ads}=10^4$ $s^{-1}$ we observed more than one Ag atom diffusing
simultaneously on the surface, building in some cases dimers that also
diffuse.

For $v_{ads}=10^5$ $s^{-1}$ we observed the formation of islands on the
surface, initially triangular like. Domains of sites corresponding to
adsorption on fcc or hcp sites are formed. The surface becomes completely
filled at $t=7\times 10^{-5}s$

For $10^6 \leq v_{ads}/s{-1} \leq 10^8$ hcp and fcc domains are
formed on the surface (Figure 11), with the domain border fluctuating rapidly
within one near neighbor distance.

According to the simulation results for this phase, we conclude that for the
deposition rates under which the experiments are undertaken (corresponding
to $0.1 \leq v_{ads}/s{-1} \leq 1$), no islands should be formed on the
surface. Surface diffusion is fast and allows the atoms to reach neighboring
steps on the surface, where nucleation and growth occur. Although the Ag
deposition on Au(111) seems to be a rather complex process, including the
formation of a number of expanded faces, we think that these findings,
that are based on the fast diffusional behavior of Ag on Au(111), should be at
least qualitatively valid.

\section{Conclusions}

Inspired in puzzling experimental results, we have modeled the underpotential
deposition for a Frank-van der Merwe system on (111) and (100) faces by means of a 
lattice model employing a realistic potential for the metal-metal interactions.
From a thermodynamic viewpoint, the present static calculations indicate
that the overpotentials for Ag nucleation on Au will be larger for the (111)
than for the (100) face. The reason for this behaviour is the larger energy that is
required to generate the borders of the growing Ag clusters in the former
case. 

If a negligible exchange current for adatoms is assumed, kinetic
simulations by means of dynamic Monte Carlo indicate that the nucleation and
growth should take place with characteristic time of tenths of seconds in
the case of the Au(100) face and of the order of milliseconds in the case of
the Au(111) face. Therefore, if Ag is deposited at intermediate rates on both surfaces,
island growth should occur on Au(100) but not on Au(111). It must be stressed
that this behaviour is a purely kinetic effect. On thermodynamic grounds,
the growth of the 2D phase should always start at steps and not on the terraces.

It must be emphasized that these conclusions are all drawn
from a model which is free from any adjustable parameters. A desirable
improvement of the present model is to consider the electrodesorption of
adatoms in different environments that involves a model for electron
transfer in this system. Work in this direction is in progress.

\section{Acknowledgements}

We thank CONICET, CONICOR, CIC, Secyt UNC and UNS. Program BID 802/OC-AR PICT N$^o$
06-04505 for financial support. Part of {\normalsize the present
calculations were performed on a Digital workstation (UNC) and the STM experiments were
carried out with Standard nanoscope III ECSTM (UNS), both donated by the
Alexander von Humboldt Stiftung, Germany.} One of the authors (W.J.L.) would like to
thank Prof. W. Wiesbeck, Institute of Microwave Techniques and Electronics for his collaboration 
in the field of Nanotechnology and support to continue joint activities even after the 
retirement of W.J.L. Language assistance by Pompeya Falc\'{o}n is gratefully acknowledged.\newpage\

\newpage

\section{Tables}

\textbf{Table 1:}

\textbf{ a)}Size $N_{crit}$ of the 2-D critical cluster of Ag on
Au(100) for different overvoltages $\eta $, calculated from the embedded
atom method and a simulated annealing procedure.

\begin{center}
\begin{tabular}{|l|l|}
\hline
$N_{crit}$ & $\left| \eta_{2DMe} \right|/eV $ \\ \hline
313 & 10 \\ \hline
78 & 20 \\ \hline
34 & 30 \\ \hline
20 & 40 \\ \hline
9 & 60 \\ \hline
5 & 80 \\ \hline
3 & 100 \\ \hline
1 & 170 \\ \hline
\end{tabular}
\end{center}

\textbf{b) }Sizes of critical clusters estimated for nucleation of Ag on a
Ag(100) quasi perfect surface , as reported in ref. \cite{Lorenz_Libro}

\begin{center}
\begin{tabular}{|l|l|}
\hline
$N_{crit}$ & $\left| \eta_{2DMe} \right|/eV $ \\ \hline
67 & 6 \\ \hline
25 & 10 \\ \hline
\end{tabular}
\end{center}

\newpage
\textbf{Table 2}: Average times required for the completion of the monolayer 
$\tau _{mon}$ for different adsorption rates per site $v_{ads}$. These times
correspond to values averaged over 6 simulation runs.

\begin{center}
{\centering
} 
\begin{tabular}{|l|l|}
\hline
$v_{ads}/s^{-1}$ & $\tau _{mon}/s$ \\ \hline
10$^{-4}$ & 6.7$\times 10^4$ \\ \hline
10$^{-3}$ & 8.3$\times 10^3$ \\ \hline
10$^{-2}$ & 7.50$\times 10^2$ \\ \hline
10$^{-1}$ & 70 \\ \hline
1 & 7 \\ \hline
10$^1$ & 0.69 \\ \hline
10$^2$ & 0.067 \\ \hline
\end{tabular}
\end{center}

\medskip{}

\begin{center}
{\centering 
}
\end{center}

\newpage\ 

\section{Figure Captions}

\textbf{Figure 1} In situ STM images of Ag UPD in the system $\textrm{Au(100)}/5\times 10^{-3}\textrm{M}$ $\textrm{AgClO}_4$  
+ $5\times10^{-1}\textrm{M}$ $\textrm{HClO}_4$ at $T=298 \textrm{K}$. a) $\Delta E=700$ $\textrm{mV}$, b) $\Delta
E=200$ $\textrm{mV}$, c) $\Delta E=20$ $\textrm{mV}$. $I_{tip}=20$ $\textrm{nA}$.

\textbf{Figure 2} In situ STM images of Ag UPD in the system $\textrm{Au(111)}/5\times 10^{-3}\textrm{M}$ $\textrm{AgClO}_4$  
+ $5\times10^{-1}\textrm{M}$ $\textrm{HClO}_4$ at $T=298 \textrm{K}$. a) $\Delta E=700$ $\textrm{mV}$, b) $\Delta
E=278$ $\textrm{mV}$, c) $\Delta E=27$ $\textrm{mV}$. $I_{tip}=15$ $\textrm{nA}$.

\textbf{Figure 3}: a-c) Sample environments for the motion of a Ag atom on a
Au(100) surface d) potential energy as a function of the distance along the
diffusion path for the environments shown in a-c.

\textbf{Figure 4 }Gibbs energy of 2D Ag cluster formation on Au(100) as a
function of size $N$ at different overpotentials $\eta_{2DMe} $.

\textbf{Figure 5} Gibbs energy of 2D Ag cluster formation on Au(111) as a
function of size $N$ at different overpotentials $\eta_{2DMe} $.

\textbf{Figure 6}: One half of the simulation box employed to represent Ag
nucleation and growth on a Au(100) surface.

\textbf{Figure 7}: Number of islands as a function of time for a deposition
rate of 10$^{-2}s^{-1}$.

\textbf{Figure 8. }Snapshots of a simulation with the Au(100) surface, 
$v_{ads}=10^{-2}s^{-1}$. t$_1$=1.7 s, t$_2$=5.7 s , t$_3$= 19 s, t$_4$= 42 s, t$_5$=
76 s, t$_6$=101 s. Grey squares are Ag adatoms, black squares represent Au
border atoms.

\textbf{Figure 9 } Number of islands as a function of the coverage degree for
different deposition rates $v$. Each curve represents an average over
six simulation runs.

\textbf{Figure 10. }Snapshots of a simulation with the Au(111) surface $%
v_{ads}=1$0$^3s^{-1}$. Simulation times are t$_1$= 0.8 s, t$_2$=4 s, t$%
_3 $= 6.6 s , t$_4$= 13 s , t$_5$= 42 s, t$_6$=87 s. The upper border
corresponds to the \{100\}step and the lower border to the \{111\} step.

\textbf{Figure 11. }Snapshots of a simulation with the Au(111) surface $%
v_{ads}=1$0$^5s^{-1}$. Circles and diamonds denote fcc and hcp adsorption
sites respectively. Filled and empty sites are represented by filled and
empty figures respectively. The simulation time was $1.46 \times 10^{-7}$ s.

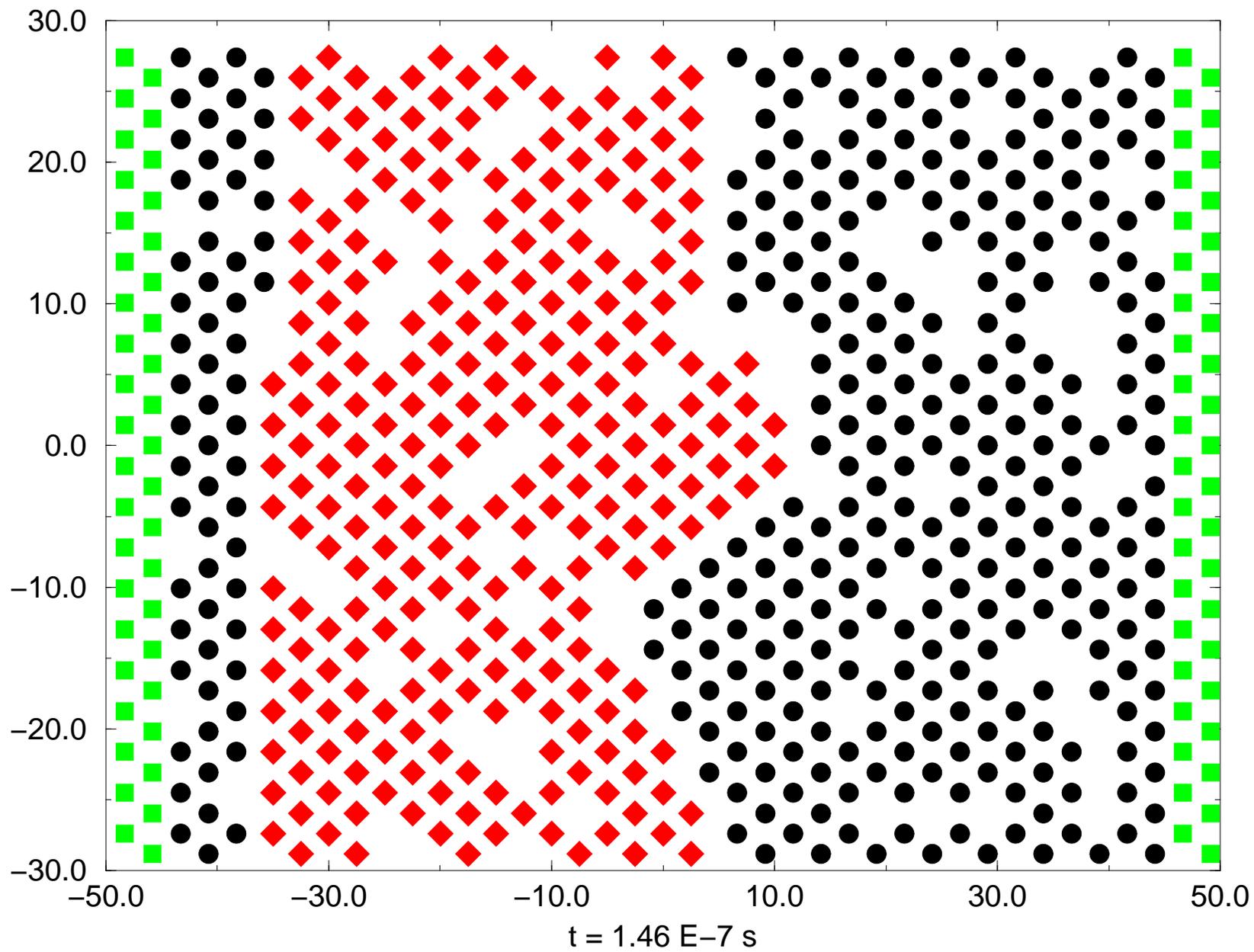
\end{document}